\documentclass[aps,prb,twocolumn,superscriptaddress,showpacs]{revtex4}
\usepackage{graphicx}
\begin{document}

\title{Angle-resolved photoemission study of quasi-one-dimensional
superconductor $\beta$-Na$_{0.33}$V$_2$O$_5$}
\author{K.~Okazaki\cite{adr}$^{,}$}
\affiliation{Department of Physics,
University of Tokyo, Bunkyo-ku, Tokyo 113-0033, Japan}
\author{A.~Fujimori}
\affiliation{Department of Physics,
University of Tokyo, Bunkyo-ku, Tokyo 113-0033, Japan}
\affiliation{Department of Complexity Science and Engineering,
University of Tokyo, Bunkyo-ku, Tokyo 113-0033, Japan}
\author{T.~Yamauchi}
\affiliation{Institute for Solid State Physics, University of Tokyo,
Kashiwanoha, Kashiwa, Chiba 277-8581, Japan}
\author{Y.~Ueda} 
\affiliation{Institute for Solid State Physics, University of Tokyo,
Kashiwanoha, Kashiwa, Chiba 277-8581, Japan}
\date{\today}

\begin{abstract}
We have studied the electronic structure of
$\beta$-Na$_{0.33}$V$_2$O$_5$, which becomes a superconductor under
pressure, by angle-resolved photoemission spectroscopy. Clear band
dispersions is observed only along the chain direction, indicating the
quasi-one-dimensional (1D) electronic structure. The spectra of the V
$3d$ band are dominated by a Gaussian-like broad feature at $\sim$ 1 eV below
the Fermi level ($E_F$) with negligible intensity at $E_F$, which we attribute
to strong electron-phonon coupling as in other 1D polaronic metals. In
the momentum space, the spectra show a maximum intensity at $k \sim \pm
\pi/4b$, where $b$ is the V-V distance along the $b$-axis,
reflecting the band filling of the chain and/or ladder.
\end{abstract}

\maketitle

\begin{figure}[t]
\includegraphics[width=6cm]{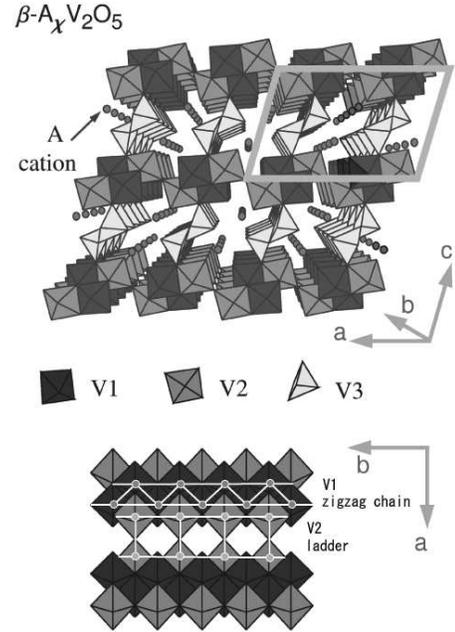} 
\caption{\label{Fig1}Crystal structure of $\beta$-vanadium bronze
 A$_{x}$V$_2$O$_5$. Schematic drawing of the V1 zigzag chain and the V2
 ladder is also shown.}
\end{figure}

The recent discovery of superconductivity ($T_{C}$ = 8 K) under high
pressure (8 GPa) in $\beta$-Na$_{0.33}$V$_2$O$_5$ attracted considerable
interest~\cite{Yamauchi} because this is the first observation of
superconductivity in vanadium oxides and because the superconductivity
occurs in a quasi-1D conductor analogous to organic superconductors.
$\beta$-Na$_{0.33}$V$_2$O$_5$ belongs to a class of $\beta$-vanadium
bronzes which have highly anisotropic crystal structures as shown in
Fig~\ref{Fig1}.  There are three kinds of V sites denoted by V1, V2,
V3. The V1O$_6$ octahedra form a zigzag chain along the $b$-axis, the
V2O$_6$ octahedra form a two-leg ladder, and the V3O$_5$ pyramids form a
zigzag chain. Previously, this compound had been studied from the view
point of a bipolaron conductor.~\cite{bipolaron} Recently, Yamada {\it
et al.}~\cite{Yamada} found that a metal-insulator transition (MIT)
occurs at $T_{MI} =$ 135 K and that above $T_{MI}$ it shows metallic
conductivity along the chain direction (the $b$-axis). Itoh {\it et
al.}~\cite{Itoh} found by an NMR study that this metal-insulator
transition is accompanied by a charge ordering below $T_{MI}$. They also
discussed the population of $d$ electrons at each vanadium site and
concluded that the V3 is V$^{5+}$-like while the V1 and/or V2 are in a
mixed-valence state. Nishimoto and Ohta studied charge ordering by
Mardelung potential calculations and found that electrons enter the V1
zigzag chain in the most stable charge
distribution~\cite{Nishimoto}. The superconductivity in this material
invoked considerable interest also because of possible roles of poralons
or electron-phonon interaction in the superconductivity. The electronic
structure is also interesting from the viewpoint of
one-dimensionality.~\cite{Haldane,Voit}

In this Rapid Communication, we report on an electronic structure study
of $\beta$-Na$_{0.33}$V$_2$O$_5$ using angle-resolved photoemission
spectroscopy (ARPES). We have found that the V $3d$ band has a
Gaussian-like broad line shape and its peak does not cross $E_F$ even in
the metallic phase. In the momentum space, on the other hand, the
maximum intensity of the V $3d$ band occurs at $k \sim \pm \pi/4b$
although in the vicinity of $E_F$, the intensity is suppressed. We shall
discuss the origins of these unusual spectral behaviors in the context of
1D metal under the strong influence of polaronic effect.

Single crystals of $\beta$-Na$_{0.33}$V$_2$O$_5$ were grown by the rf
heating Czochralski method as described in
Ref.~\onlinecite{Yamauchi}. Photoemission measurements were performed
using a Gammadata-Scienta SES-100 electron analyzer and a
Gammadata-Scienta VUV-5000 He discharge lamp with a toroidal grating
monochromator. The energy calibration and the estimation of the energy
resolution were made by measuring gold spectra. The energy resolution
and the momentum resolution were set at $\sim$ 25 meV and $\pm$
0.2$^\circ$ near $E_F$ and $\sim$ 40 meV and $\pm$ 0.4$^\circ$
otherwise. The base pressure of the analyzer chamber was 5 $\times$
10$^{-10}$ Torr. Clean surfaces were obtained by cleaving the sample
{\it in situ}.  The reproducibility was checked by repeated cleaving and
measurements.


\begin{figure}[t]
\begin{center}
\includegraphics[width=7cm]{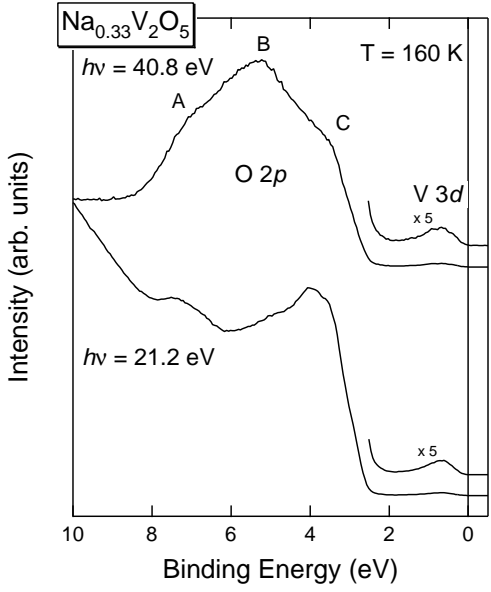}
\caption{\label{Fig2}Normal emission spectra of
$\beta$-Na$_{0.33}$V$_2$O$_5$ taken at 160 K.}
\end{center}
\end{figure}
Figure~\ref{Fig2} shows the spectra in the entire valence-band region
taken with He {\footnotesize I} ($h\nu$ = 21.2 eV) and He {\footnotesize
II} ($h\nu$ = 40.8 eV) at normal emission. The structures at 2.5-8 eV
binding energies ($E_B$'s) are originated from the O $2p$ band while the
weak feature at $E_B$ $\simeq$ 1 eV is originated from the V $3d$
states. Because there is only one V $3d$ electron per six vanadium atoms,
the intensity of the V $3d$ band is extremely weak. Among the three
structures in the O $2p$ band, A, B and C, structure B is enhanced in
the He {\footnotesize II} spectrum compared to He {\footnotesize I},
indicating that this structure is strongly hybridized with the V $3d$
states since the relative photoionization cross section of the V $3d$
states to the O $2p$ states is larger for the He {\footnotesize II}
spectra than for the He {\footnotesize I} spectra.~\cite{Yeh}
\begin{figure}[t]
\begin{center}
\includegraphics[width=5.5cm]{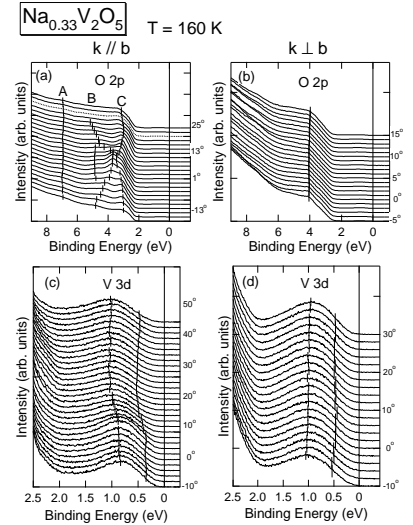}
\caption{\label{Fig3}ARPES spectra of
$\beta$-Na$_{0.33}$V$_2$O$_5$ taken at 160 K. The vertical lines
indicate peak positions or inflection points.}
\end{center}
\end{figure}

Figure~\ref{Fig3} shows the angular dependence of the spectra taken at
160 K. The left panels (a) and (c) show the spectra with momentum along
the chain ($b$-axis) while the right panels (b) and (d) show the spectra
with momentum perpendicular to the chain (i. e., along the
$c$-axis). The vertical lines in the O $2p$ band indicate local minimum
positions in the second-derivative of the spectra while those in the V
$3d$ band indicate peak positions and the inflection points determined
by the first and second derivatives, respectively. Clear dispersion can
be seen only along the chain direction both for the O $2p$ and the V
$3d$ bands. Structure B is the most strongly dispersive in the O $2p$
band probably because this structure is most strongly hybridized with
the V $3d$ states as mentioned above. The V $3d$ band has a broad line
shape and the peak position is located at $E_B$ $\sim$ 0.9 eV. The
dispersional width is $\sim$ 0.25 eV.

The peak position of the V $3d$ band in $\beta$-Na$_{0.33}$V$_2$O$_5$ is
closer to $E_F$ than that in the related insulating vanadates
NaV$_2$O$_5$ ($E_B$ $\sim$ 1.5 eV, Ref.~\onlinecite{Kobayashi1}) and
LiV$_2$O$_5$ ($E_B$ $\sim 1.3$ eV, Ref.~\onlinecite{LiV2O5}) with $d^1$
electronic configuration, reflecting the smaller band filling or the
lower $E_F$ in the present compound. However, the intensity is
suppressed near $E_F$ and the peak does not disperse sufficiently to
allow the band crossing $E_F$. The ARPES spectra of a quasi-1D conductor
(TaSe$_4$)$_2$I, which shows a CDW transition at 160 K, as reported by
Perfetti {\it et al}~\cite{Perfetti1} also show a similar behavior in
its metallic phase and only the tail of the Ta $5d$ band reaches $E_F$.
They interpreted this characteristic spectral feature of (TaSe$_4$)$_2$I
in terms of a polaron picture~\cite{phonon} because electron-phonon
interactions are obviously important in CDW systems such as
(TaSe$_4$)$_2$I. In fact, for $\beta$-Na$_{0.33}$V$_2$O$_5$ strong
electron-phonon coupling had been proposed.~\cite{bipolaron} The
Gaussian-like broad feature of the V $3d$ band and the small spectral
weight at $E_F$ may therefore be attributed to strong electron-phonon
coupling. The high resistivity ($\sim 10^{-2} \Omega {\rm cm}$) for a
metal may indicate that electron-phonon interaction is very strong and
that the charge carriers are mobile small polarons.~\cite{Perfetti2}
\begin{figure*}[t]
\begin{center}
\includegraphics[height=7cm]{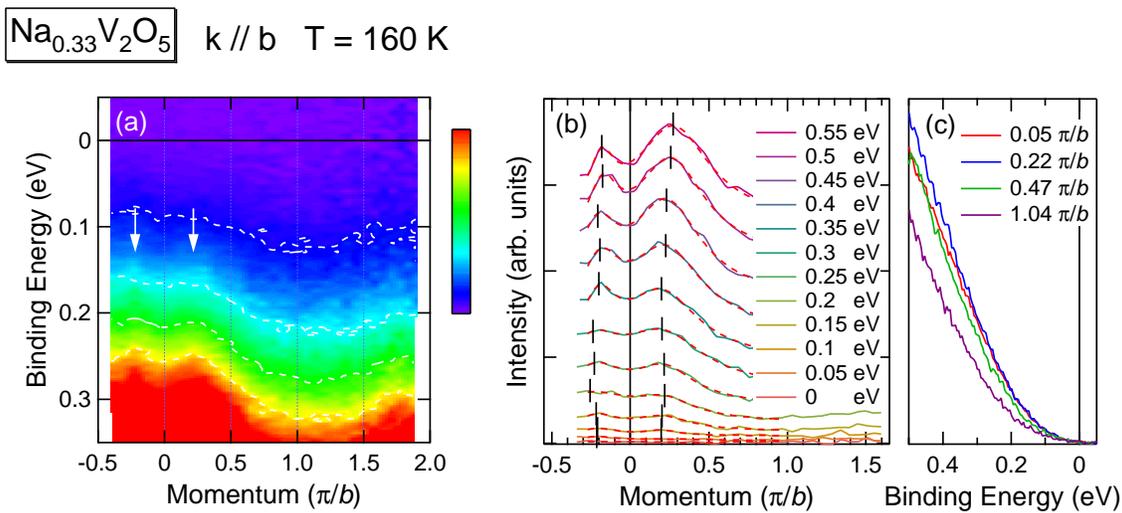} 
\caption{\label{Fig4} (a) Intensity plot of ARPES spectra near $E_F$ in
 the $E$-$k$ plane. The vertical arrows indicate $k = \pm \pi/4b$. (b)
 Momentum distribution curves. Dashed curves are fitted Lorentzians and
 the vertical lines indicate their peak positions. (c) Energy
 distribution curves.} 
\end{center}
\end{figure*}

The intensity plot of the ARPES spectra near $E_F$ in the $E$-$k$ plane
is shown in Fig.~\ref{Fig4} (a). One can see that the contour lines
become closest to the $E_F$ at $k \sim \pm \pi/4b$. From the fitting of
the momentum distribution curves (MDCs) to two Lorentzians, as shown in
Fig.~\ref{Fig4} (b), their peak positions indicated by the vertical bars
are $\pm$(0.22 $\pm$ 0.02)$\pi/b$ ($\sim \pm\pi/4b$), where $b$ is the
V-V distance along the $b$-axis of the zigzag chains or of the ladder
(see Fig.~\ref{Fig1} (b)), and depend on $E_B$ only very weakly
(corresponding to the Fermi velocity of $>$ 10 eV$\cdot$\AA). This means
that although most of the spectral weight of the V $3d$ band is confined
in the Gaussian-like broad peak centered at $E_B$ $\sim$ 0.9 eV and the
Gaussian peak does not cross $E_F$, there are underlying strongly
dispersive features and their Fermi surface crossings at $k \simeq \pm
\pi/4b$. Strongly dispersive weak feature crossing $E_F$ has been
observed in a double-layer metallic manganite
La$_{1.2}$Sr$_{1.8}$Mn$_2$O$_7$ (Refs.~\onlinecite{Mn1,Mn2}), where the
spectra are also dominated by a broad Gaussian-like peak at higher
binding energy and a very weak feature reaches and cross $E_F$.  To
explain this unusual spectral behavior, a polaron model has been
proposed.~\cite{Mn1,Mn2} 

The occurrence of the maximum intensities at $k \simeq \pm \pi/4b$ is
contrasted to the corresponding feature in the insulating NaV$_2$O$_5$
(Ref.~\onlinecite{Kobayashi1}) and LiV$_2$O$_5$
(Ref.~\onlinecite{LiV2O5}), whose band maximum and hence the maximum
intensity near $E_F$ occurs at $k \simeq \pm \pi/2b$. The smaller value
$k \simeq \pm \pi/4b$ in the present compound should be related to the
smaller band filling of the chains and/or ladders in
$\beta$-Na$_{0.33}$V$_2$O$_5$. Based on this observation, in the
following we discuss possible band filling of the V1 zigzag chain and
the V2 ladder for several cases depending on the relative strength of
the nearest-neighbor and next-nearest-neighbor V-V hopping in the chain
and the relative strength of the V-V hopping within the rung and within
the leg of the ladder. First, we consider the case where the bonding
within the rung and the nearest-neighbor hopping in the zigzag chain are
strong. If the doped electrons enter the V $3d$ orbitals only of the
ladder, the ladder would become quarter-filled (1/2 electron per V2
site) and the band filling would be similar to NaV$_2$O$_5$. In the high
temperature phase of NaV$_2$O$_5$, because the bonding orbital of the
rung is occupied and the antibonding orbital is empty, the
quarter-filled ladder effectively behaves as a half-filled 1D chain, and
the ARPES spectra are well reproduced by the half-filled 1D {\it t-J}
model or Hubbard model~\cite{Kobayashi2}, resulting in the band maximum
at $k = \pm \pi/2b$. We expect the same situation for the quarter-filled
ladder in $\beta$-Na$_{0.33}$V$_2$O$_5$. If the doped electrons enter
only the zigzag chain, the chain would be quarter-filled and the
Fermi-level crossing (and hence the intensity maximum) would occur again
at $k = \pm \pi/2b$. (Note that there are two V1 atoms per $b$.) If
$\sim$ 50\% of doped electrons go into the zigzag chain (resulting in
the 1/8-filled zigzag chain) and the remaining $\sim$ 50\% go into the
ladder (the 1/8-filled ladder), Fermi-level crossing would occur at $k =
\pm \pi/4b$, consistent with experiment. In this case, the band filling
of the chain and ladder is determined by a subtle balance between the
energy levels of the chain and the ladder and therefore may vary
sensitively under high pressure. Electrons may be transferred between
the zigzag chain and the ladder under high pressure as in the hole-doped
ladder system Sr$_{14-x}$Ca$_{x}$Cu$_{24}$O$_{41}$
(Ref.~\onlinecite{14-24-41}), which also shows a pressure-induced
superconductivity.

Second, we consider the case where the bonding within the rung of the
ladder is weak or the nearest-neighbor hopping in the zigzag chain is
weak compared to the next-nearest-neighbor hopping in the chain. Indeed,
the bonding within the rung of the ladder in
$\beta$-Na$_{0.33}$V$_2$O$_5$ may be weaker than that of NaV$_2$O$_5$,
because the V-V distance is longer: 3.6 {\AA} for
$\beta$-Na$_{0.33}$V$_2$O$_5$ {\it versus} 3.45 {\AA} for
NaV$_2$O$_5$. In that case, the ladder or the zigzag chain can be
regarded as two independent single chains. Therefore, if $\sim$ 50\% of
doped electrons go into one of the two independent chains and the
remaining $\sim$ 50\% go into another chain, these chains become
quarter-filled single chains and the intensity maximum would occur at $k
= \pm \pi/4b$ near $E_F$. Hence, electrons may enter both of the ladder
and the zigzag chain, or may enter either the ladder or the chain,
because that ladder or zigzag chain consists of two quarter-filled
chains. The existence of quarter-filled single chains is analogous to
the quarter-filled quasi-1D band in organic superconductors such as
Bechgaard salts.~\cite{Bechgaard}

Since $\beta$-Na$_{0.33}$V$_2$O$_5$ is shown to be a 1D metal above
$T_{MI}$ (= 135 K), we have attempted to analyze the photoemission
spectra within the Tomonaga-Luttinger (TL) liquid frame work. In
Fig.~\ref{Fig4} (c), energy distribution curves (EDCs) near $E_F$ are
shown for the metallic phase. As mentioned above, the intensity is
suppressed near $E_F$, which is a characteristic of a TL liquid. The
line shape is approximately parabolic for all momenta. Therefore, the
anomalous exponent $\alpha$ is at least $\sim$ 2 in the present
system,~\cite{Voit,note} extremely large compared to other 1D conductors
($\alpha \lesssim 1$).~\cite{Denlinger,Xue,Mizokawa1} Note that the
Hubbard model predicts $\alpha < 1/8$ (Ref.~\onlinecite{Schulz}). One
possible origin of the large $\alpha$ is that electron-phonon
interaction is strong compared to other systems. Another possibility is
that the long-range Coulomb interaction is unscreened because there are
few conducting carriers. Theoretically, it has been shown that if
long-range interaction exists, the exponent $\alpha$ can be any large
value.~\cite{Kawakami}

Finally, we briefly comment on the temperature dependence of the spectra
across $T_{MI}$. The leading edge of the V $3d$ band is shifted toward
higher binding energies by $\sim$ 15 meV in going from the metallic to
insulating phases, taken at 160 K and 100 K, respectively. We have also
studied changes in the band dispersion from the peak positions and the
inflection points [Fig.~\ref{Fig3} (c)].  We could not, however, observe
clear change from those plots across $T_{MI}$.


In conclusion, we have studied the electronic structure of
$\beta$-Na$_{0.33}$V$_2$O$_5$. The band dispersion is finite only along
the chain direction, indicating the quasi-1D electronic structure. In
spite of the metallic conductivity, the V $3d$ band has a Gaussian-like
broad feature whose peak does not cross the Fermi level. We attributed
this to the strong electron-phonon coupling in the quasi-1D metal. In
the momentum space, the V $3d$ band shows maximum intensities at $k \sim
\pm \pi/4b$, due to a Fermi-level crossing or a valence-band
maximum. Based on this observation, we have discussed possible band
filling of the zigzag chain and the ladder in
$\beta$-Na$_{0.33}$V$_2$O$_5$.

Discussions with T. Ohta and S. Nishimoto are gratefully
acknowledged. This work was supported by a Grant-in-Aid for Scientific
Research in Priority Area ``Novel Quantum Phenomena in Transition
Metal Oxides'' from the Ministry of Education, Culture, Sports, Science
and Technology, Japan.


\begin{thebibliography}{99}

\bibitem[*]{adr}Present address: Institute for Solid State Physics,
 University of Tokyo, Kashiwanoha, Kashiwa, Chiba 277-8581, Japan

\bibitem{Yamauchi}
T. Yamauchi, Y. Ueda, and N. M\^{o}ri, Phys. Rev. Lett. {\bf 89},
057002 (2002).

\bibitem{bipolaron} B. K. Chakraverty, M. J. Sienko, and J. Bonnerot,
Phys. Rev. B {\bf 17}, 3781 (1978).

\bibitem{Yamada}
H. Yamada and Y. Ueda, J. Phys. Soc. Jpn. {\bf 68}, 2735 (1999).

\bibitem{Itoh}
M. Itoh, N. Akimoto, H. Yamada, M. Isobe, and Y. Ueda, J. Phys.
Soc. Jpn. {\bf 69}, Suppl. B 155 (2000).

\bibitem{Nishimoto}
S. Nishimoto and Y. Ohta, J. Phys. Soc. Jpn. {\bf 70}, 309 (2001).

\bibitem{Haldane}
F. D. M. Haldane, J. Phys. C, {\bf 14,}, 2585 (1981)

\bibitem{Voit}
J. Voit, Rep. Prog. Phys., {\bf 58}, 977 (1995).

\bibitem{Yeh}
J. -J. Yeh and I. Lindau, At. Data Nucl. Data Tables {\bf 32}, 1
(1985).

\bibitem{Kobayashi1}
K. Kobayashi, T. Mizokawa, A. Fujimori, M. Isobe, and Y. Ueda,
Phys. Rev. Lett. {\bf 80}, 3121 (1998).

\bibitem{LiV2O5}
K. Okazaki, A. Fujimori, M. Isobe, and Y. Ueda, unpublished.

\bibitem{Kobayashi2}
K. Kobayashi, T. Mizokawa and A. Fujimori and M. Isobe, Y.
Ueda, T. Tohyama, and S. Maekawa, Phys. Rev. Lett. {\bf 82}, 803
(1999).

\bibitem{Perfetti1}
L. Perfetti, H. Berger, A. Reginelli, L. Degiorgi, H. H\"{o}chst,
J. Voit, G. Margaritondo, and M. Grioni, Phys. Rev. Lett. {\bf
87}, 216404 (2001).

\bibitem{phonon}
G. D. Mahan, {\it Many-Particle Physics} (Plenum, New York, 1981)
Sec. 4.3.

\bibitem{Perfetti2}
L. Perfetti, S. Mitrovic, G. Margaritondo, M. Grioni, L.
Forr\'{o}, L. Degiorgi, and H. H\"{o}chst, Phys. Rev. B {\bf 66},
075107 (2002).

\bibitem{Mn1}
D. S. Dessau, T. Saitoh, C. -H. Park, Z. -X. Shen, P. Villella,
	N. Hamada, Y. Moritomo, and Y. Tokura, Phys. Rev. Lett. {\bf
	81}, 192 (1998).

\bibitem{Mn2}
Y. -D. Chuang, A. D. Gromko, D. S. Dessau, T. Kimura, and Y. Tokura,
	Science {\bf 292}, 1509 (2001).
\bibitem{Schulz}
H. J. Schulz, Phys. Rev. Lett. {\bf 64}, 2831 (1990).

\bibitem{Bechgaard}
e.g. C. Bourbonnais, and D. J\'{e}rome, {\it Advances in Synthetic
	Metals, Twenty Years of Progress in Science and Technology},
	eds. by P. Bernier, S. Lefrant, and G. Bidan (Elsevier, New
	York, 1999).

\bibitem{14-24-41}
N. Motoyama, T. Osafune, T. Kakeshita, H. Eisaki, and S. Uchida,
Phys. Rev. B, {\bf 55}, R3386 (1997).

\bibitem{Kawakami}
N. Kawakami and S. -K. Yang, Phys. Rev. Lett,{\bf 67}, 2493
(1994).

\bibitem{note}
We have estimated $\alpha$ by fitting the angle-integrated spectra from
$E_F$ to $E_B = 0.1$ with a power-law function $A\omega^{\alpha}$. The
estimated $\alpha$ is $1.7 \pm 0.1$. 

\bibitem{Denlinger}
J. D. Denlinger, G. -H. Gweon, J. W. Allen,  C. G. Olson, J.
Marcus, C. Schlenker, and L. -S. Hsu, Phys. Rev. Lett, {\bf 82},
2540 (1999).

\bibitem{Xue}
J. Xue, L. -C. Duda, and K. E. Smith, A. V. Fedorov, P.D. Johnson, S. L. 
Hulbert, W. McCarroll, M. Greenblatt, Phys. Rev. Lett. {\bf
83}, 1235 (1999).

\bibitem{Mizokawa1}
T. Mizokawa, K. Nakada, C. Kim, Z. -X. Shen, T. Yoshida, A.
Fujimori, S. Horii, Yuh Yamada, H. Ikuta, and U. Mizutani, Phys.
Rev. B, {\bf 65}, 193101, (2002).

\end{thebibliography}
\end{document}